\documentstyle[12pt]{article}
\input epsf

\newcommand{\eps}{\epsilon}
\newcommand{\tw}{\theta_{\!\mathrm{w}\,}}
\newcommand{\sintw}{\sin\tw}
\newcommand{\costw}{\cos\tw}
\newcommand{\Fem}{{\cal F}^{\mathrm{em}}}
\newcommand{\Fmn}[1]{{\cal F}_{\mu\nu}^{(#1)}}
\newcommand{\Aem}{A^{\mathrm{em}}}
\newcommand{\phat}{\hat{\phi}}
\newcommand{\trace}{\mathop{\mathrm{Tr}}}
\newcommand{\e}{{\mathrm{e}}}
\renewcommand{\d}{{\mathrm{d}}}
\renewcommand{\i}{{\mathrm{i}}}
\newcommand{\bm}[1]{{\mbox{\boldmath $#1$}}}
\newcommand{\qed}{~~$\Box$}
\newcommand{\Cset}{{\rm C}\kern-2.3mm\raisebox{.7mm}{$\scriptstyle
\mid$}\kern2mm}
\renewcommand{\AA}{{\bf A}\kern-1.5mm\raisebox{.7mm}{$\scriptstyle
\backslash$}}

\def\np#1#2#3{{\it  Nucl.\ Phys.\ }{{\bf #1}\ {(#2)}\ {#3}}}
\def\pr#1#2#3{{\it Phys.\ Rev.\ }{{\bf #1}\ {(#2)}\ {#3}}}
\def\pl#1#2#3{{\it  Phys.\ Lett.\ }{{\bf #1}\ {(#2)}\ {#3}}}

\def\ijmp#1#2#3{{\it Int.\ J.\ Mod.\ Phys.\ }{{\bf #1}\ {(#2)}\ {#3}}}

\begin{document}
\footnotesep=16pt
\begin{flushright}
\baselineskip=14pt
{\normalsize DAMTP-1998-33}\\
{\normalsize {hep-ph/9805255}}
\end{flushright}
\renewcommand{\thefootnote}{\fnsymbol{footnote}}
\baselineskip=24pt

\begin{center}
{\Large\bf Definition of the Electromagnetic Field}\\
{\Large\bf in the Broken-Symmetry Phase}\\
{\Large\bf of the Electroweak Theory}\\
\baselineskip=16pt
\vspace{0.75cm}

{\bf Ola T\"{o}rnkvist}\footnote{\baselineskip=16pt Electronic
address: {\tt O.Tornkvist@damtp.cam.ac.uk}}\\
\vspace*{4mm}
{\em Department of Applied Mathematics and Theoretical Physics}\\
{\em University of Cambridge, Cambridge CB3 9EW, United Kingdom}
\vspace*{0.75cm}

May 6, 1998
\end{center}
\baselineskip=20pt
\begin{quote}
\begin{center}
{\bf\large Abstract}
\end{center}
\vspace{0.2cm}
{\baselineskip=10pt
In the broken-symmetry phase of the electroweak
theory there is no unique definition of
the electromagnetic field tensor in cases where
the magnitude of the Higgs field differs from a constant value.
The meaning of the electromagnetic field is therefore
dubious near defects and during
non-equilibrium stages of the electroweak phase transition.
Nevertheless, by imposing a minimal set of natural requirements one is
led to a
specific, gauge-invariant definition
that retains the familiar properties of an electromagnetic field.
An electromagnetic vector potential is
constructed whose curl (exterior derivative)
in any gauge gives the electromagnetic field
tensor. As is required, this vector potential
transforms at most by a pure gradient under arbitrary
$\mathrm{SU(2)}\times\mathrm{U(1)}$ gauge transformations.
The flux of the magnetic field is expressed
as a gauge-invariant line integral. Curiously, this
provides a definition for
magnetic flux in cases where the spatial region with broken symmetry
is not simply connected and the magnetic field itself is not everywhere
defined.}
\vspace*{8pt}
\noindent

\end{quote}
\renewcommand{\thefootnote}{\arabic{footnote}}
\setcounter{footnote}{0}
\newpage
\baselineskip=20pt

\section{Introduction}

It is well-known that the concept of an electromagnetic field
has no meaning in the symmetric phase of the electroweak theory,
since vector potentials may there be rotated into each other
by gauge transformations.
In contrast,
one may
expect
that the electromagnetic field
should become
uniquely defined
as soon as the
SU(1)${}_{\mathrm L}\times$U(1)${}_Y$ symmetry
breaks to U(1)${}_{\mathrm EM}$,
because
the photon field is then distinguished
as the only vector field with
zero electric charge and zero mass.

Despite this anticipated uniqueness, several
gauge-invariant definitions of the electromagnetic field tensor
are in common use. It was recently discovered by this author
\cite{myPRD} that even those definitions that coincide
when the Higgs magnitude
$\rho=(\Phi^{\dagger}\Phi)^{1/2}$ is constant, give
different results
when $\rho$ has a space-time dependence.
Unless this ambiguity can be resolved,
statements about the strength, presence or
absence of electromagnetic fields are meaningless when characterising
field configurations that include a variation of the magnitude
of the Higgs field, e.g.
near \mbox{(non-)topological} defects,
on the walls of expanding bubbles of the broken-symmetry phase in a
first-order electroweak phase transition, or in the hot
early universe subject to large thermal fluctuations of the Higgs
field.

In the Higgs ground state, characterised by a constant Higgs field with
magnitude $\rho\equiv v$, the electromagnetic fields of everyday life
are distinguished by a set of properties that uniquely set them apart
from other
strong and electroweak interactions:
\begin{itemize}\parsep=0pt\parskip=0pt\itemsep=0pt
\item[A.] An electromagnetic field is a long-range field, i.e.\
it always extends from point sources or line sources according to
a power law without exponential suppression.
\item[B.] There are no magnetic charges or magnetic currents that can
generate an electromagnetic field.
\item[C.] An electromagnetic field is never generated by an electrically
neutral current.
\end{itemize}

In this Letter we show
that the definition of the electromagnetic field tensor
 can be extended to a
general gauge and space-time varying Higgs magnitude $\rho(x)>0$ in such
a way that all the above intuitive aspects of electromagnetic fields are
preserved.

In section 2 we investigate the properties and
physical consequences of various proposed
gauge-invariant definitions of the electromagnetic field tensor in the
electroweak theory. By imposing the long-range force requirement
(property A) one finds that all the preexisting definitions
are eliminated except the tensor $\Fem_{\mu\nu}$
proposed in Ref.~\cite{myPRD}, given here in eq.~(\ref{Fdef}).

This argument
does not prove {\em per se\/}
the non-existence of other
definitions that might fulfil the requirement. It does, however,
point to a field tensor which has the property
C and which satisfies
the Bianchi identity,
implying not only that property B is accommodated, but also
that the field tensor may be written in any gauge as the curl of a
vector potential.

This vector potential is constructed in
section 3 and in Appendix A. First, it is noted that
the electromagnetic vector potential
in the unitary gauge is given by the usual massless field $A_\mu$,
also when $\rho$ has a space-time dependence. By applying
a general gauge transformation to the vector fields, and expressing
the SU(2) part of
this gauge transformation in terms of the Higgs field,
a vector potential $\Aem_\mu$
is constructed
with the property that it changes at most by a gradient under
arbitrary SU(2)$\times$U(1) transformations and reduces to $A_\mu$ in
the unitary gauge. Taking the curl of $\Aem_\mu$ is shown to produce
$\Fem_{\mu\nu}$.

Finally, in section 4 a gauge-invariant
definition of magnetic flux is obtained as the line integral
of the electromagnetic vector potential $\Aem_i$ along a closed curve.
Curiously, this provides an expression for the magnetic flux
also in cases where the spatial region
with broken symmetry is not simply connected and the magnetic
field itself is not everywhere defined.

\section{Gauge-Invariant Definitions
of the\\ Electromagnetic Field Tensor}
\label{GIFsec}

The mass eigenstates of the vector-boson fields are determined by
the Lagrangian kinetic term for the Higgs field,
$({\cal D}^\mu\Phi)^{\dagger} {\cal D}_\mu \Phi$, where
${\cal D}_\mu = \partial_\mu -\i (g/2)W^a_\mu\tau^a - \i(g'/2)Y_\mu$.
With conventional definitions $A_\mu=\sintw W^3_\mu + \costw
Y_\mu$, $Z_\mu = \costw W^3_\mu - \sintw Y_\mu$, and
$W_\mu = (W^1_\mu - \i W^2_\mu)/\sqrt{2}$ we can write the covariant
derivative ${\cal D}_\mu\Phi$ as
\begin{equation}
{\cal D}_\mu\Phi = \left(\!\!
\begin{array}{lr}
\partial_\mu -\i e A_\mu - \i \gamma Z_\mu &
\hspace{-2em} -\i\frac{M_W}{v} W_\mu \\*[1ex]
-\i \frac{M_W}{v} W_\mu^{\dagger} &\hspace{-2em}
\partial_\mu + \frac{\i}{\sqrt{2}}
\frac{M_Z}{v} Z_\mu
\end{array}\!\!
\right)
\Phi~,
\label{covder}
\end{equation}
where $\tan\tw = g'/g$, $2\gamma= g\cos 2\tw/\costw$, $e=$ $g\sintw$,
$M_W^2 = g^2 v^2/2$ and $M_Z=M_W/\costw$.

In the unitary gauge the Higgs field is given by
\begin{equation}
\Phi=\left(\begin{array}{c} 0\\* \rho(x)\end{array}\right)~,\quad
\rho(x)\geq 0~,
\label{ugauge}
\end{equation}
and it follows
immediately from eq.~(\ref{covder})
that only the field $A_\mu$ is
massless, while the fields $W_\mu$ and $Z_\mu$ acquire masses.
What is usually not mentioned in textbooks is
that this holds true regardless
of the functional form of the magnitude $\rho(x)$. Therefore,
in a unitary
gauge given by eq.~(\ref{ugauge}), $A_\mu$ is
always the electromagnetic vector potential.

Although global field configurations may sometimes be expressed
in the unitary gauge,
this is usually not possible when the Higgs field $\Phi$ has zeros.
For example,
in vortex solutions such as the electroweak $W$-string or $Z$-string
\cite{embdef},
the winding of
the Higgs isospin orientation is such that the vector fields
become singular when transformed to the unitary gauge.
Another example concerns the collision of expanding bubbles of the
broken-symmetry phase in a first-order electroweak phase transition.
In general the bubbles, having had no previous causal contact, will
contain Higgs fields with different isospin orientation, and it would
be awkward to describe them in a unitary gauge. Worse still,
since field configurations similar to electroweak strings
can be produced in such bubble collisions
\cite{Copeland2}, imposing the unitary gauge would
result in a loss of generality.

Consider therefore a general gauge with
Higgs field $\Phi=(\varphi_1(x), \varphi_2(x))^{\top}$. The
field $A_\mu$ defined in eq.~(\ref{covder})
then couples to $\varphi_1(x)$ and
becomes massive. Evidently, at positions $x$ where $\varphi_1(x)\neq
0$, $A_\mu$ is no longer the photon field.
Consequently, the electromagnetic field tensor $\Fem_{\mu\nu}$
can no longer have
the same expression as in the unitary gauge.
Instead,
one must construct a gauge-invariant definition of $\Fem_{\mu\nu}$ whose
value in any gauge coincides
(except at points where $\Phi=0$) with that obtained by (locally)
transforming all fields to the unitary gauge and evaluating
it there.
To this end, let us define a three-component unit isovector
$\phat^a=(\Phi^{\dagger}\tau^a\Phi)/(\Phi^{\dagger}\Phi)$, where
$\tau^a$, $a=1\ldots 3$, are the Pauli spin matrices. Note that $\phat^a$
is independent of the magnitude $\rho$ and depends only on the
isospin orientation of $\Phi$.

Let us start by considering a gauge-invariant
definition first proposed by Nambu
\cite{Nambu} and subsequently
used in investigations of the distribution of
electromagnetic fields and charges inside the electroweak sphaleron
\cite{Hindmarsh,sphal}. It is given by
\begin{equation}
{\cal F}_{\mu\nu}^{\mathrm N} :=-\sintw \phat^a F^a_{\mu\nu} +
\costw F^Y_{\mu\nu}~,
\label{Nambudef}
\end{equation}
where
$F^a_{\mu\nu} = \partial_\mu W^a_\nu
- \partial_\nu W^a_\mu +
g\eps^{abc} W^b_\mu W^c_\nu$
and $F^Y_{\mu\nu}=\partial_\mu Y_\nu - \partial_\nu Y_\mu$ are
the SU(2)$_{\mathrm L}$ and
U(1)$_Y$ field tensors, respectively. In the unitary gauge
(\ref{ugauge}) we have
$\phat^a=-\delta^{a3}$
and ${\cal F}_{\mu\nu}^{\mathrm N}$ reduces to
\begin{equation}
{\cal F}_{\mu\nu}^{\mathrm N}=
A_{\mu\nu} + \i e (W^{\dagger}_\mu W_\nu - W^{\dagger}_\nu W_\mu)
\label{unambu}
\end{equation}
with
$$A_{\mu\nu}\equiv \partial_\mu A_\nu - \partial_\nu A_\mu~.$$

A peculiar, and in our opinion unattractive, property of this
definition is that it admits finite-range electromagnetic fields.
This
becomes apparent in the unitary gauge
if we consider a field configuration with
$A_{\mu\nu}=0$
containing
a localised distribution of charged $W$ bosons with
$W^{\dagger}_{[\mu} W_{\nu]}\neq 0$. Although this is not a
common-day occurrence, such a configuration of $W$ fields is
in principle realisable in nature.
Because the $W$ fields are massive, the electromagnetic fields given
by ${\cal F}^{\mathrm N}_{\mu\nu}$ are non-zero but
decay exponentially away from their
sources.
This
marks a departure from the usual, intuitive notion of
electromagnetic fields as being long-range fields with a power-law
behaviour. Moreover, since $A_\mu$ is a massless field in the unitary
gauge regardless of $\rho(x)$,
we are guaranteed of the existence of another
field tensor, e.g.\ $A_{\mu\nu}$,
which always decays away from its sources according to a power
law, and if this is not the electromagnetic field, one may have
to consider a different name for the field with this property.

There is a clarifying analogy with Maxwell's equations in a
medium which corroborates this picture. We can identify the field
tensor components ${\cal F}_{ij}^{\mathrm N}$ with the {\em magnetic
intensity\/} $\bm{H}$, whose sources are only the {\em free\/}, or
external, currents. The magnetic field, or {\em induction\/}, $\bm{B}$
is in the unitary gauge
identified with $A_{ij}$ and is produced by all currents,
including the magnetisation currents in the medium. These fields
are related by
$\bm{H}=\bm{B}-4\pi \bm{M}$, where the magnetisation $\bm{M}$
can be identified with the $W$-boson terms.

To justify this analogy,
let us consider a uniform
magnetic intensity $H=|\bm{H}|$ in the $x_3$ direction.
If the external current is sufficiently high,
so that $H$ exceeds
the critical value $M_W^2/e$, the electroweak vacuum becomes
unstable with respect to the production of a
condensate of $W$-boson pairs
\cite{Ambjorn} in the spin polarisation state $W_1=-\i W_2\equiv W$.
With $H={\cal F}_{12}^{\mathrm N}$ and $B=A_{12}$
it follows that $B=H+2e|W|^2$, and one finds that
the $W$ bosons contribute a positive
magnetisation $M=e |W|^2/(2\pi)$. As a consequence
the vacuum is paramagnetic for $H>M_W^2/e$, as was shown in
Ref.~\cite{Ambjorn}.

Let us return to the Gedanken experiment
with the localised $W$-boson distribution. It follows from
the field equations for $A_{\mu\nu}$ that, in order to obtain
$A_{\mu\nu}=0$, which was part of the premises, one would have to
screen the current of the $W$ fields by means of an external current.
If we now interpret $A_{ij}$ as the magnetic field, the mutual screening
of the two currents provides a {\em physical\/} reason for $A_{ij}$
to be zero. Moreover, there is no contradiction between  $A_{ij}$
being zero at large distances and being a long-range force with
a power-law behaviour, because
$A_{ij}$ is zero everywhere.

Armed with this new intuition, let us now investigate
some alternative gauge-invariant
definitions of the electromagnetic field
tensor.

In order to
obtain the full electromagnetic field $A_{\mu\nu}$ in the unitary
gauge, i.e.\ not only the
part generated by free currents, one
would have to
subtract the $W$-field terms of ${\cal F}^{\mathrm N}_{\mu\nu}$
in a gauge-invariant way.
This problem was
partly solved by Vachaspati,
who proposed the following field tensor \cite{Vacha},
\begin{equation}
{\cal F}^{\mathrm V}_{\mu\nu}:=
 - \sintw{\phat}^a F^{a}_{\mu\nu}  +
\costw F^{Y}_{\mu\nu}
- \i\frac{\sintw}{g}\frac{2}{\Phi^{\dagger}\Phi}
\left[({\cal D}_\mu\Phi)^{\dagger} {\cal D}_\nu\Phi -
({\cal D}_\nu\Phi)^{\dagger} {\cal D}_\mu\Phi\right]~.
\label{Vexp}
\end{equation}

The added term in the above expression
cancels the quadratic terms in the $W$ field correctly when
$\rho$ is constant, but introduces
extraneous terms when $\rho$ has a space-time dependence. This becomes
apparent
in the unitary gauge, where
${\cal F}^{\mathrm V}_{\mu\nu}$
reduces to \cite{Grasso}
\begin{equation}
{\cal F}^{{\mathrm V}}_{\mu\nu} = A_{\mu\nu} - 2 \tan\tw
(Z_\mu\partial_\nu \ln\rho - Z_\nu\partial_\mu \ln\rho)~.
\label{Uvacha}
\end{equation}
Consider now the case $A_{ij}=0$ with non-zero field components
$Z_i$ and
gradient of the Higgs magnitude $\partial_i\rho$,
such that $\eps_{ijk}Z_j\partial_k \rho\neq 0$.  This situation is
characteristic of the interior of the electroweak $Z$-string solution
\cite{embdef}. With the definition (\ref{Uvacha}) one would then
conclude that there is a magnetic field present \cite{Grasso}.
However, as is shown in detail in Ref.~\cite{myPRD},
this definition would imply that electromagnetic fields can be
generated by electrically neutral currents. These currents are
obtained in the unitary gauge by taking the divergence of
eq.~(\ref{Uvacha}), and consist of derivatives of the neutral fields
$Z_\mu$ and $\rho$.

Because of the cylindrical symmetry of the $Z$-string solution,
even a long-range
electromagnetic field
with power-law behaviour away from point
or line sources
would be zero outside the string
(cf.\ a solenoid).
For $A_{\mu\nu}=0$,
and for
$Z_{[\mu}\partial_{\nu]} \rho\neq 0$ inside some finite volume
${\cal V}$ with less symmetry,
one can show as before that the
electromagnetic fields given by
${\cal F}^{{\mathrm V}}_{\mu\nu}$ are
non-zero within ${\cal V}$, but decay exponentially away from
${\cal V}$,
thereby departing from the expected behaviour of electromagnetic
fields away from sources of generic geometry.

A definition of the electromagnetic field tensor
that exhibits none of the above unattractive features was proposed
in Ref.~\cite{myPRD}:
\begin{equation}
\Fem_{\mu\nu} \equiv -
\sintw{\phat}^a F^{a}_{\mu\nu}  +
\costw F^{Y}_{\mu\nu} +
\frac {\sintw}{g}  \epsilon^{abc}
{\phat}^{a} (D_\mu\phat)^{b} (D_\nu\phat)^{c}~,
\label{Fdef}
\end{equation}
where $(D_\mu\phat{})^a=\partial_\mu\phat^a + g\epsilon^{abc}
W^b_\mu\phat^c$.
It reduces to $A_{\mu\nu}$ in the unitary gauge
(\ref{ugauge})
for arbitrary
$\rho=\rho(x)$.
The fact that $A_\mu$ is always a massless field in this gauge
ensures that $\Fem_{\mu\nu}$ is a long-range field with a
power-law behaviour.
In fact, $\Fem_{\mu\nu}$
is the {\em unique\/} gauge-invariant realisation of $A_{\mu\nu}$,
since two gauge-invariant tensors that agree in one gauge have
the same value in any gauge. The definition (\ref{Fdef}) has appeared
previously in Ref.~\cite{Hindmarsh} and, for the Glashow-Georgi
SO(3) model, in Ref.~\cite{tHooft}.

Unlike ${\cal F}^{\mathrm N}_{\mu\nu}$ and ${\cal F}^{\mathrm
V}_{\mu\nu}$,
the electromagnetic field tensor $\Fem_{\mu\nu}$ satisfies
the Bianchi identity
\mbox{${\cal F}^{\rm em}_{[\mu\nu,\alpha]}=0$}
everywhere except on the worldlines of magnetic monopoles.
Therefore, in the absence of monopoles, there
is no magnetic charge or magnetic current.
Furthermore, there are no contributions to $\Fem_{\mu\nu}$ from
electrically neutral currents. The electric current
$j^{\mathrm e}_\nu\equiv\partial^{\mu}\Fem_{\mu\nu}$ and its properties
were discussed at length in Ref.~\cite{myPRD}.

An important consequence of the definition (\ref{Fdef})
is that the
electromagnetic field tensor
receives contributions
from gradients of the phases (i.e.\ isospin orientation) of the
Higgs field, even when the vector potentials are zero.
To better understand this aspect,
consider a field configuration where all the vector potentials
$W^a_\mu$ and $Y_\mu$ are zero, but the Higgs
field is of the general form
\begin{equation}
\Phi=\rho\left(
\begin{array}{c}
\e^{\i\alpha}\sin\omega\\*
\e^{-\i\beta}\cos\omega
\end{array}
 \right)~,
\label{Hangles1}
\end{equation}
where $\rho$, $\alpha$, $\beta$, and $\omega$ are all functions of the
space-time coordinates $x^\mu$.
This configuration can be transformed to the unitary gauge by means
of a gauge transformation $\Phi\to U\Phi$ with $U\in\mathrm{SU(2)}$
defined by
\begin{equation}
U=\left(\begin{array}{cc}
\e^{-\i\beta}\cos\omega & -\e^{\i\alpha}\sin\omega\\*
\e^{-\i\alpha}\sin\omega & \e^{\i\beta}\cos\omega
\end{array}\right)~.
\label{Udef}
\end{equation}
We then obtain, in the unitary gauge, a Lie-algebra valued vector
potential ${\AA}_\mu = -i(\partial_\mu U) U^{\dagger}$ with
components $W^a_\mu=\trace(\tau^a {\AA}_\mu)/g$ and
$Y_\mu=\trace {\AA}_\mu/g'$. In this gauge the vector potential
$A_\mu=\sintw W^3_\mu + \costw Y_\mu$ becomes
\begin{equation}
A_\mu=\frac{2\sintw}{g} \left[
\sin^2\!\omega\,\partial_\mu \alpha -
\cos^2\!\omega\,\partial_\mu \beta\right]~.
\label{Aemreg1}
\end{equation}
and its curl gives the electromagnetic field
\begin{equation}
A_{\mu\nu}=\frac{2\sintw}{g}\sin 2\omega
(\omega_{[,\mu}\alpha_{,\nu]} +\omega_{[,\mu}\beta_{,\nu]})~.
\label{ubfield}
\end{equation}
On the other hand, one can evaluate the
field tensor $\Fem_{\mu\nu}$ of eq.~(\ref{Fdef})
directly for
the general Higgs configuration (\ref{Hangles1}) and
$W_\mu^a=Y_\mu=0$. The last term of $\Fem_{\mu\nu}$ then takes into
account the electromagnetic field associated with the space-time
dependent Higgs isospin orientation, and reproduces
the right-hand side of eq.~(\ref{ubfield}).
Such an electromagnetic
field was not included in
definition
(\ref{Nambudef}).

It is interesting to note that the expression (\ref{Nambudef})
was derived by Nambu
as the most general non-zero field configuration
compatible with the two conditions $\rho\equiv v$ and
${\cal D}_\mu \Phi=0$, which together
define the Higgs vacuum. The purpose of his defi\-nition was
to characterise
the behaviour of fields {\em far\/}
from defects, not in the interior of
defects where these two conditions are violated.
Indeed, in the unitary gauge the condition ${\cal D}_\mu \Phi=0$
implies that $W_\mu=W^{\dagger}_\mu=0$, so if definition (\ref{Nambudef})
is used
only in the restricted case
${\cal D}_\mu \Phi=0$
as was intended,
the field tensor
${\cal F}^{\mathrm N}_{\mu\nu}$ reduces to $A_{\mu\nu}$ in the
unitary gauge.

\section{``Gauge-Invariant'' \mbox{Vector}
Poten\-tial for the Electro\-magnetic
Field}
\label{GIAsec}

Because the preferred
electromagnetic field tensor (\ref{Fdef}) satisfies the
Bianchi identity \linebreak[4]
\mbox{$\Fem_{[\mu\nu,\alpha]}=0$}, it is
possible to construct a vector potential $\Aem_{\mu}$
for the electromagnetic
field with the property that $\Fem_{\mu\nu}=\partial_\mu \Aem_{\nu} -
\partial_\nu \Aem_{\mu}$.
Since $\Fem_{\mu\nu}$ is gauge-invariant,
the vector potential
must transform at most by a pure gradient under arbitrary
$\mathrm{SU(2)}\times \mathrm{U(1)}$ gauge transformations.
Appendix A describes the method for constructing vector potentials
with this invariance. Here we shall state only the results.

A non-zero scalar-doublet Higgs field $\Phi=(\varphi_1,\varphi_2)^{\top}$
can always be written
$\Phi = V (0,\rho)^{\top}$, where
$\rho = (\Phi^{\dagger}\Phi)^{1/2}$
is the magnitude and
the isospin orientation is defined by the SU(2)-valued function
$V$,
\begin{equation}
V=\frac{1}{\rho} \left(
\!\!\begin{array}{cc}
(\i\tau^2\Phi)^{\ast}&\Phi \end{array}\!\!\right)=
\frac{1}{\rho}\left(\!\!
\begin{array}{rc}
\varphi_2^{\ast}&\varphi_1\\*
-\varphi_1^{\ast}&\varphi_2 \end{array}\!\!\right)~.
\label{Vdef}
\end{equation}
The vector potential $\Aem_{\mu}$ can then be expressed as
\begin{equation}
\Aem_{\mu}=\sintw\!\!\left\{-\phat^a W^a_\mu +\frac{\i}{g}
\trace (\tau^3 V^{\dagger}\partial_\mu V)\right\}
+\costw Y_\mu~.
\label{Adef1}
\end{equation}
As is shown in Appendix A, this expression for $\Aem_{\mu}$ is in
fact invariant under SU(2) gauge transformations, and changes by
a pure gradient under U(1) transformations.

A more practical expression, for which the connection to
eq.~(\ref{Fdef}) is easier to establish, is obtained by
rewriting
$V$ in terms of the unit isovector $\bm{\phat}=\{\phat^a\}
\in S^2$ and a
U(1) phase.
One finds
\begin{equation}
V = \left(\!\!\begin{array}{cc}
1-\phat^3 & \phat^1-\i \phat^2\\*
-\phat^1-\i \phat^2 & 1 - \phat^3
\end{array}\!\!\right)
\frac{\e^{-\i\xi\tau^3}}{\sqrt{2(1-\phat^3)}}~,
\label{Vdef2}
\end{equation}
where $\xi$ is the phase of the lower component of
$\Phi$.
By means of the unit-vector constraint $\phat^a\phat^a=1$
the vector potential $\Aem_{\mu}$ can now be written
\begin{equation}
\Aem_{\mu}=-\sintw\phat^a W^a_\mu +\costw Y_\mu
-\frac{\sintw}{g}\frac{1}{1-\phat^3}
\eps^{3ab}\phat^a\partial_\mu\phat^b~.
\label{Adef2}
\end{equation}
A term
$\partial_\mu(2\sintw\xi/g)$ was omitted
here,
as $\Aem_{\mu}$ is
defined only up to a gradient.

In eq.~(\ref{Adef2}) the
apparent singularity of $\Aem_{\mu}$ at $\phat^3=1$, corresponding
to Higgs fields of the form $\Phi=(\phi,0)^{\top}$, $\phi\in\Cset$,
is merely an artefact of the coordinate system chosen in
eq.~(\ref{Vdef2}) and
arises because there is no global decomposition of SU(2) as a Cartesian
product of $S^2$ and U(1).
If one instead
parametrises the isospin orientation of the
Higgs field
by angles $\alpha$, $\beta$ and $\omega$
as in eq.~(\ref{Hangles1}),
then the last term of eq.~(\ref{Adef2}) becomes
\begin{equation}
\frac{2\sintw}{g} \sin^2\!\omega\, \partial_\mu(\alpha+\beta)~,
\label{Aemreg}
\end{equation}
which is regular everywhere.

Although eqs.~(\ref{Adef2}) and (\ref{Fdef}) look superficially
similar,
the proof that
 $\Fem_{\mu\nu}=$\linebreak[4] $\partial_\mu \Aem_{\nu} -
\partial_\nu \Aem_{\mu}$ requires some effort, and is deferred to
Appendix B. Instead, let us here establish that $\Aem_{\mu}$
changes at most by a pure gradient under arbitrary gauge
transformations.

First, under a U(1) transformation $\Phi\to \e^{\i\theta}\Phi$,
the only change is $Y_\mu\to Y_\mu + 2\partial_\mu\theta/g'$, which
gives a pure gradient in $\Aem_{\mu}$. Continuing with
the group SU(2)
we consider, for simplicity, infinitesimal transformations
defined by
\begin{eqnarray}
\Phi&\to& (1 + \frac{\i}{2}\omega^a(x) \tau^a)\Phi~,\nonumber\\*
\phat^a&\to&\phat^a - \eps^{abc}\omega^b\phat^c~,\nonumber\\*
W_\mu^a&\to& W_\mu^a - \eps^{abc}\omega^bW_\mu^c +
\frac{1}{g}\partial_\mu\omega^a~.
\label{SU2trans}
\end{eqnarray}
Inserting this into eq.~(\ref{Adef2}) and expanding in $\omega^a$,
the linear terms are
\begin{eqnarray}
&&{\displaystyle
\frac{\sintw}{g}\left\{\frac{1}{1-\phat^3}
[\omega^3_{,\mu} -\phat^a
\omega^a_{,\mu}
- \omega^a\phat^a
\phat^3_{,\mu} +
\phat^3\omega^a
\phat^a_{,\mu}]
+\frac{1}{(1-\phat^3)^2}
\eps^{3ab}\eps^{3cd}\phat^a\phat^b_{,\mu}\omega^c\phat^d\right\}}
\nonumber\\*
&&\quad\quad=
{\displaystyle\partial_\mu\left[\frac{\sintw}{g(1-\phat^3)} (\omega^3 -
\phat^a\omega^a)\right]~.}
\label{linom}
\end{eqnarray}
In the last step, leading again to a pure gradient, the
constraint $\phat^a\phat^a=1$ was used.

In the unitary gauge, where $\Phi=(0,\rho)^{\top}$ and
$\phat^a\equiv-\delta^{a3}$, the electromagnetic
potential $\Aem_{\mu}$ reduces to the usual expression
$A_\mu = \sintw W^3_\mu + \costw Y_\mu$.
This fact, together with the invariance property of
$\Aem_{\mu}$ under gauge transformations,
independently establishes that $\Fem_{\mu\nu}=
\partial_\mu\Aem_{\nu} - \partial_\nu\Aem_{\mu}$.

It is noteworthy that, although expressions similar to
eq.~(\ref{Adef1}) have occurred previously \cite{Vacha,Hindmarsh},
the crucial term with the matrix $V$ was missing. This term
is needed to cancel the inhomogeneous Maurer-Cartan term
acquired by the vector potentials under gauge transformations.
Expressions without this term are
not gauge-invariant and also fail
in general to project out the massless component of the gauge
potential.

\section{Gauge-Invariant Definition of Magnetic Flux}
\label{GIflux}
Because of the existence of a vector potential $\Aem$
such that $\Fem=\d\wedge\Aem$, we have by Stokes' theorem
that
\begin{equation}
\int_S \Fem_{\mu\nu}\ \d x^\mu\!\wedge \d x^\nu =
\oint_{\partial S} \Aem_{\mu} \d x^\mu~,
\label{Stokes}
\end{equation}
where $S$ is an oriented surface with
correspondingly oriented boundary $\partial S$.
Therefore, the flux $\Phi_B$ of the magnetic field $B^i=\frac{1}{2}\eps^{ijk}
\Fem_{jk}$ across a surface S with normal $\bm{n}$ is given by
\begin{equation}
\Phi_B = \int_S B^i n^i \d S = \oint_{\partial S} \Aem_{i} \d x^i~.
\label{bflux}
\end{equation}
This expression is invariant under SU(2)$\times$U(1) gauge
transformations and
is useful for evaluating the magnetic
flux in practical applications such as  computer simulations of
the production of magnetic
fields in bubble collisions in the
electroweak phase
transition.
Numerical evaluation of the flux then requires only one-dimensional
integration.

Let us now consider a rather curious example where the line integral
of $\Aem_{i}$ along the curve $\partial S$ has a well-defined value, but
the magnetic field $B^i$ is ill-defined on a subset of the surface $S$.
In such a case, Stokes' theorem cannot strictly be said to hold, but
the line integral in eq.~(\ref{bflux}) can nevertheless be used to
define a magnetic flux.

\begin{figure}[htb]
\epsfxsize=9cm
\begin{center}
\leavevmode
\epsfbox{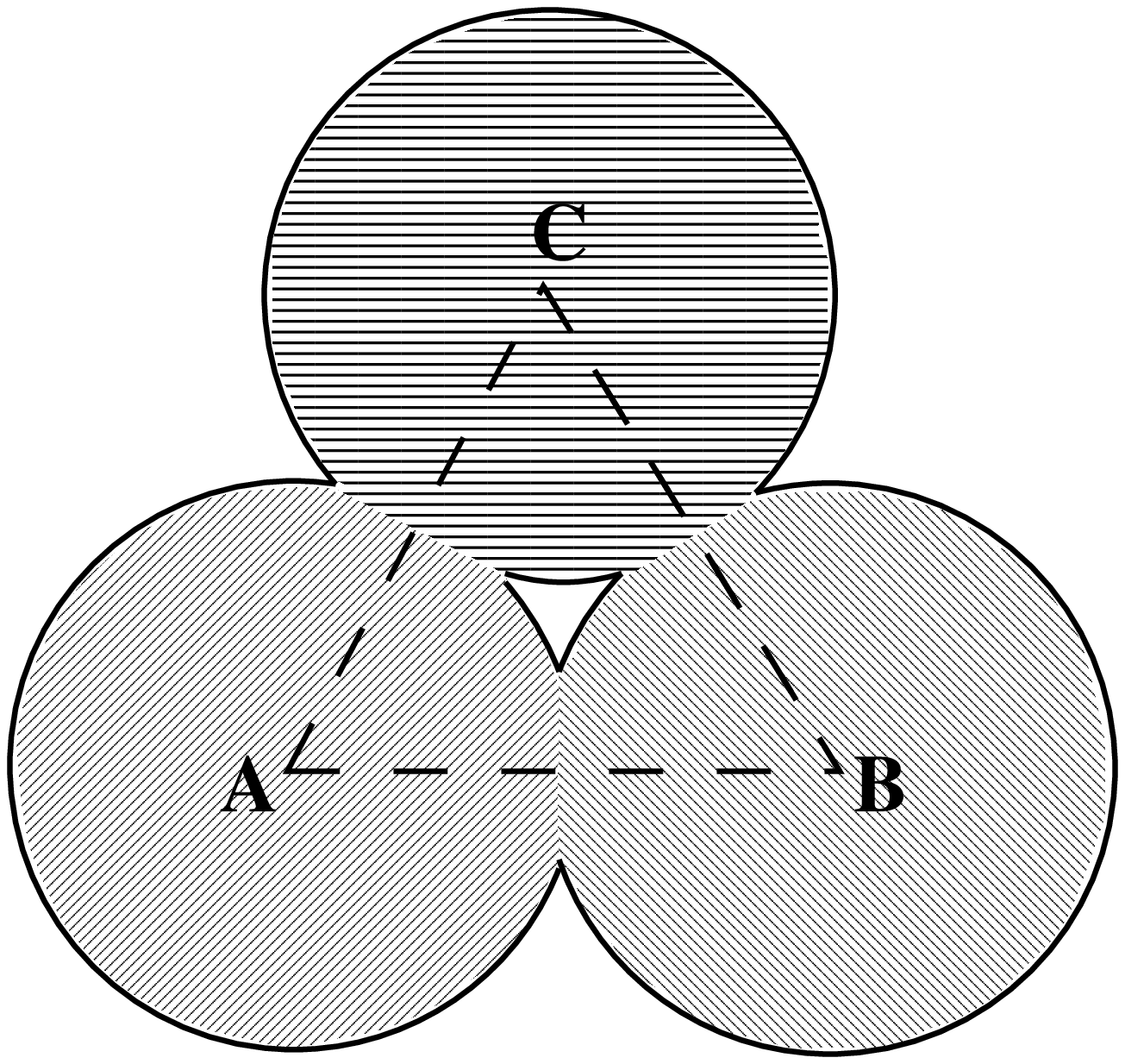}
\end{center}
\caption{\baselineskip=16pt%
Collision of three expanding bubbles in the electroweak
phase transition. The shaded region corresponds to the broken-symmetry
phase.}
\label{bubfig}
\end{figure}

Figure 1 shows a section in the plane of collision of
three expanding spherical
bubbles of the broken-symmetry phase in a first-order
electroweak phase transition. If the Higgs field
has different isospin orientation in each of the three bubbles,
it can be shown
that a magnetic field is produced in the collision
\cite{Ambleside,prep}. This magnetic
field is initially given by the last term in eq.~(\ref{Fdef}).
At the particular instant depicted, the bubbles
have recently joined at the midpoints of the line segments $\mathrm{AB}$,
$\mathrm{BC}$ and $\mathrm{CA}$,
but a triangle-like region at the center of the
figure has not yet been reached by the expanding bubble walls and
still contains the symmetric phase with $\Phi=0$. In this region,
$\phat^a$ and the magnetic field $\Fem_{ij}$ are
undefined.\footnote{\footnotesep=16pt%
In cases where the
contour $\partial S$ encloses one or more string defects, the set where
$\Phi=0$ and $\Fem_{ij}$ is
ill-defined
has measure zero, and both integrals in eq.~(\ref{bflux}) are
well-defined.}
Nevertheless, since $\Aem_{i}$ is defined everywhere along the
contour $\mathrm{ABCA}$, the magnetic flux through the triangle bounded by
this contour can be defined as
\begin{equation}
\Phi_B =  \oint_{{\mathrm{ABCA}}} \Aem_{i} \d x^i~.
\label{triflux}
\end{equation}
Except in pathological gauges, the components $\Aem_i$ are continuous
functions of time along the path ${\mathrm ABCA}$. Therefore, the
expression (\ref{triflux}) will approach continuously the value
of the surface integral of the magnetic field
at the instant when the hole between the bubbles
disappears and the region of broken symmetry becomes simply connected
so that eq.~(\ref{bflux}) becomes valid.

\section{Conclusions}
We have shown that the definition of the electromagnetic field tensor
can be extended to a general gauge and to a space-time varying Higgs
magnitude $\rho(x)$ in such a way that the familiar properties of
an electromagnetic field are retained. More precisely, the
gauge-invariant field tensor $\Fem_{\mu\nu}$ defined by
eq.~(\ref{Fdef}) is a long-range field with power-law behaviour
away from point sources
or line sources (A), satisfies the Bianchi identity
everywhere except on the worldlines of magnetic monopoles (B),
and is never generated
by an electrically neutral current (C).

These three properties do not uniquely define an electromagnetic field
tensor. Even in the Higgs ground state with
$\rho\equiv v$,
the choice $\Fem_{\mu\nu}=A_{\mu\nu}$ (unitary gauge),
with $A_{\mu\nu} =
\partial_\mu A_\nu - \partial_\nu A_\mu$,
is a matter of long-standing convention.
One may add, for example, $\partial_{[\mu}(A_{\nu]\alpha} h^\alpha)$ to the
field tensor, where $h^\alpha$ is any tensor containing charged
fields that is invariant under the
unbroken U(1) symmetry, without affecting any of the properties A, B
or C. Nonetheless, the conventional
choice $A_{\mu\nu}$ is the simplest field tensor that
can be constructed from $A_\mu$, and the tensor $\Fem_{\mu\nu}$
defined
by eq.~(\ref{Fdef})
is its unique
gauge-invariant extension.

The field tensor $\Fem_{\mu\nu}$ has the important property that it
receives contributions from gradients of the phases of the Higgs field
also when the vector potentials are zero. This is as
should be expected,
because under a transformation to the unitary gauge these
gradients are converted into non-zero
vector potentials that contribute to $A_{\mu\nu}$. The generation
of electromagnetic fields from such Higgs gradients has important
applications in cosmology, e.g.\ in electroweak bubble collisions
\cite{Copeland2,Ambleside,prep}, and may likewise have some
significance in the
interior of defects such as the electroweak sphaleron
\cite{Hindmarsh,sphal}.

Finally, we have constructed a ``gauge-invariant'' vector potential
$\Aem_{\mu}$
for the electromagnetic field with the property that it transforms
at most by a pure gradient under arbitrary SU(2)$\times$U(1)
gauge transformations. The field tensor $\Fem_{\mu\nu}$
is given in any gauge by the curl of $\Aem_{\mu}$, which by
Stokes' theorem implies that magnetic flux can be expressed as
a  gauge-invariant line integral.

\subsection*{Acknowledgments}
I am grateful to Paul Saffin for useful comments and for checking several
of the results. This work was supported by EPSRC under grant GR/K50641.

\section*{Appendix A: SU(2)-invariant vector potentials}

In this appendix we describe the method for
constructing vector
potentials invariant under SU(2) gauge transformations. In
the unitary gauge, where $\Phi=\Phi_0\equiv (0,\rho)^{\top}$,
let the Lie-algebra valued vector potential be given by
$\bar{\AA}_\mu
= (g/2) \bar{W}^a_\mu\tau^a + (g'/2)
\bar{Y}_\mu$.
The vector potential in an arbitrary gauge with
Higgs field $\Phi=V\Phi_0$ is
then found by means of an SU(2)$\times$U(1) gauge
transformation defined by the function $U=V\exp(\i\lambda Q)$, where
$V$ is given by eq.~(\ref{Vdef}) and the charge operator
$Q=(1+\tau^3)/2$
satisfies $Q\Phi_0=0$. Under this transformation
$\bar{{\AA}}_\mu\to{{\AA}}_\mu = U\bar{{\AA}}_\mu
U^\dagger - \i(\partial_\mu U) U^{\dagger}$.
Solving for $\bar{{\AA}}_\mu$, we obtain
\begin{equation}
\bar{{\AA}}_\mu =\e^{-\i\lambda Q}[V^{\dagger}{\AA}_\mu
V + \i V^{\dagger}\partial_\mu V]\e^{\i\lambda Q} -
\partial_\mu\lambda Q~.
\label{newold}
\end{equation}

Because $\bar{{\AA}}_\mu$ is the
vector potential in the unitary gauge, it is, by definition,
gauge-invariant, and so are its components $\bar{W}^3_\mu =
\trace(\tau^3 \bar{{\AA}}_\mu)/g$ and $\bar{Y}_\mu
= \trace  \bar{{\AA}}_\mu/g'$.
Now write the
vector potential in the new, arbitrary gauge as
${{\AA}}_\mu= (g/2) W^a_\mu\tau^a + (g'/2) Y_\mu$.
Using the identity $V\tau^3 V^{\dagger} = -\phat^a\tau^a$
one obtains
\begin{eqnarray}
\bar{W}^3_\mu&=&-\phat^a W^a_\mu + \frac{\i}{g}\trace(\tau^3
V^{\dagger}\partial_\mu V) - \frac{1}{g}\partial_\mu\lambda~,
\nonumber\\*
\bar{Y}_\mu&=&Y_\mu - \frac{1}{g'}\partial_\mu\lambda~.
\label{gipot}
\end{eqnarray}
The gauge-invariant electromagnetic vector potential is
given by its value in the unitary gauge,
$\bar{A}_\mu =\sintw \bar{W}^3_\mu + \costw \bar{Y}_\mu$.
This differs from the expression (\ref{Adef1}) only by a
pure-gradient term $-\partial_\mu\lambda/e$, which may be omitted.
Whereas $V$ is uniquely determined by the Higgs field
$\Phi$, the function
$\lambda(x)$ is arbitrary and corresponds to the unbroken
U(1) symmetry of electromagnetism.

\section*{Appendix B: Curl of the electromagnetic vector
potential}

In this appendix we prove that the curl of $\Aem_{\mu}$, given by
eq.~(\ref{Adef2}), equals $\Fem_{\mu\nu}$, expressed in
eq.~(\ref{Fdef}). The constraint $\phat^a\phat^a=1$ will be employed
repeatedly. We start by expanding
$\Fem_{\mu\nu}$,
noting that terms quadratic in the
fields $W_\mu^a$ cancel, and obtain
a sum of four terms,
\begin{eqnarray}
\Fem_{\mu\nu}=-\sintw\phat^a\partial_{[\mu}W^a_{\nu]}+
\costw F^Y_{\mu\nu}\quad\quad~\nonumber\\*
- \sintw W^a_{[\nu} \partial_{\mu]} \phat^a
+\frac{\sintw}{g}\eps^{abc}\phat^a\partial_\mu\phat^b\partial_\nu
\phat^c~.
\label{Fdef2}
\end{eqnarray}
Let us denote the terms of eq.~(\ref{Adef2}) by $A_{~\mu}^{(i)}$,
$i=1\ldots 3$, and those of
eq.~(\ref{Fdef2}) by $\Fmn{i}$,
$i=1\ldots 4$. Taking the curl of the first term of $\Aem_{\mu}$,
one obtains $\partial_{[\mu}A_{~\nu]}^{(1)} =
 \Fmn{1}+\Fmn{3}$.
Trivially, $\partial_{[\mu}A_{~\nu]}^{(2)} =
\Fmn{2}$. It remains
to show that $\partial_{[\mu}A_{~\nu]}^{(3)} = \Fmn{4}$.
For simplicity set $\sintw/g=1$, as it is a common factor, and
consider
\begin{equation}
\partial_{[\mu}A_{~\nu]}^{(3)}=-\frac{2}{1-\phat^3}
\eps^{3bc}\partial_\mu\phat^b\partial_\nu\phat^c
-
\frac{1}{(1-\phat^3)^2}\partial_{[\mu}\phat^3\eps^{3bc}\phat^b
\partial_{\nu]}\phat^c~.
\label{A3def}
\end{equation}
By means of the identity $\partial_{[\mu}\phat^3\eps^{3bc}\phat^b
\partial_{\nu]}\phat^c= -\eps^{abc}\phat^a\partial_\mu\phat^b
\partial_\nu\phat^c + \eps^{3bc}\phat^3\partial_\mu\phat^b
\partial_\nu\phat^c$ the right-hand side can be rearranged to give
\begin{equation}
\partial_{[\mu}A_{~\nu]}^{(3)}=\frac{1}{(1-\phat^3)^2}
\left[
\eps^{abc}\phat^a\partial_\mu\phat^b
\partial_\nu\phat^c + (\phat^3-2)
\eps^{3bc}\partial_\mu\phat^b
\partial_\nu\phat^c
\right]~.
\label{A3step}
\end{equation}
The identity
$\eps^{3bc}\partial_\mu\phat^b
\partial_\nu\phat^c$ $=$ $\phat^3\eps^{abc}\phat^a\partial_\mu\phat^b
\partial_\nu\phat^c$ leads to the result
$\partial_{[\mu}A_{~\nu]}^{(3)}$ $=$ $\eps^{abc}\phat^a\partial_\mu\phat^b
\partial_\nu\phat^c = \Fmn{4}$,
which completes the proof. \qed

%




\end{document}